\documentclass{IEEEtran}
\usepackage{amstext}
\usepackage{cite}
\usepackage{amsmath}
\usepackage{amsbsy}
\usepackage{amssymb}
\usepackage{mathtools}
\usepackage{graphicx}
\usepackage[font={small},labelfont={bf}]{caption}
\usepackage{subcaption}
\usepackage{multirow}
\usepackage{booktabs}
\usepackage{flushend}
\usepackage{soul}
\usepackage{pifont}

\usepackage[UKenglish]{babel}
\usepackage{graphicx}
\usepackage[usenames,dvipsnames]{color}
\usepackage{ifthen}
\usepackage{algorithm}
\usepackage[]{algorithmic}
\usepackage{enumitem}
\usepackage{lipsum}

\newcommand{\normdist}[2]{\mathcal{N} \brackets{{#1,#2} }    }
\newcommand{\normpdf}[3]{\mathcal{N} \brackets{{#1; #2, #3} }    }

\newcommand{\beq}{\begin{eqnarray}}
\newcommand{\eeq}{\end{eqnarray}}
\newcommand{\inmat}[1]{\begin{bmatrix}#1\end{bmatrix}}
\newcommand{\eqn}[2]{\begin{align}\ifthenelse{\equal{#1}{}}{\nonumber}{} #2\ifthenelse{\equal{#1}{}}{\nonumber}{\label{eq:#1}}\end{align}}
\newcommand{\brackets}[1]{\left({#1}\right)}

\newcommand{\Algo}[1]{Algorithm \ref{alg:#1}}

\begin{document}

\title{ If and When a Driver or Passenger is Returning to Vehicle: Framework to Infer Intent and Arrival Time }
\author{Bashar I. Ahmad, 
        Patrick M. Langdon,
        Simon J. Godsill,
        Mauricio Delgado 
         and 
        Thomas Popham
 \thanks{ B. I. Ahmad and S. J. Godsill are with the Signal Processing and Communications (SigProC) Laboratory, Engineering Department, University  of Cambridge, Trumpington Street, Cambridge, UK, CB2 1PZ. Emails:\{bia23, sjg30\}@cam.ac.uk.}
\thanks{P. M. Langdon is with the Engineering Design Centre (EDC), Engineering Department, University of Cambridge, Trumpington Street, Cambridge, UK, CB2 1PZ. Email:pml24@cam.ac.uk.}
 \thanks{M. Delgado and T. Popham are with Jaguar Land Rover,  Coventry, UK. Emails: \{tpopham, amunozd1\}@jaguarlandrover.com.}
}


\maketitle


\begin{abstract}
This paper proposes a probabilistic framework for the sequential estimation  of the likelihood of a driver or passenger(s) returning to the vehicle and time of arrival, from the available partial track of the user's location. The latter can be provided by a smartphone navigational service and/or other dedicated (e.g. RF based)  user-to-vehicle positioning solution. The introduced novel approach treats the tackled problem as an intent prediction task within a Bayesian formulation, leading  to an efficient implementation of the inference routine with notably low training requirements. It  effectively captures the long term dependencies in   the trajectory followed by the driver/passenger to the vehicle, as dictated by intent, via a bridging distribution.
 Two examples are shown to demonstrate the efficacy of this  flexible low-complexity technique. 
\end{abstract}
\IEEEpeerreviewmaketitle

\section{Introduction}
\subsection{Background and Motivation}
The recent advances in sensing, data storage as well as processing and communications technologies led to the proliferation of intelligent vehicle functionalities and services. This includes Advanced Driver Assistance Systems (ADAS)\cite{bengler2014three,paul2016advanced}, route guidance \cite{gwon2016generation,Gustafsson2017}, driver inattention monitoring \cite{dong2011driver} and many others  \cite{bishop2005intelligent,fleming2016early}.  Additionally, whilst   the current growing interest in autonomous cars brings a myriad of new technical and human factor challenges \cite{neumann2016automated,greene2016our}, it has encouraged expediting the development and adoption of smart  vehicle   services and their associated technologies. In particular, there has been a phenomenal growth of research into realising a  connected cooperative vehicle environment \cite{coppola2016connected,zheng2015heterogeneous,Schwarz2017,seredynski2016survey}, which is key to the success of autonomous driving as well as  enhancing transportation efficiency and safety. This encompasses vehicle to vehicle, vehicle to infrastructure, vehicle to devices and vehicle to cloud communications, imposing new  requirements on in-vehicle systems and the supporting infrastructure. 


Within the context of intelligent vehicles, perhaps in a connected set-up, there are substantial benefits to be gained from determining if and when the driver or
passenger(s) is returning to  vehicle, as early as possible and before the start of a journey.
For instance, it can enable the: 
\begin{enumerate}
\item
 timely adaption of the car interior to \textit{a priori} learnt preferences or driver/passenger(s) profiles (e.g. adjusting seats and pre-configuring the infotainment system, adapting the  HMI, warming/cooling vehicle, etc.); thereby delivering a personalised, safer and more pleasant driving experience,
\item efficient  activation and/or priming of the key-fob scanner (e.g. for key-less entry or engine start)  and exterior-facing vehicle sensors (e.g.  cameras for driver/passenger recognition), which can   also improve their security features,

\end{enumerate}
to name a few. 

 In this paper, we address the problem of establishing the intent of  a driver or passenger (i.e. whether returning to car) and estimating time of arrival from his/her available partial location trajectory,  possibly in a connected vehicle environment. This track   can be provided by the user's smartphone Global Navigation Satellite System (GNSS) service    or a dedicated user-to-vehicle positioning solution.
\vspace{-4.5mm}
\subsection{Contributions}
The problem of determining if and when a driver/passenger is returning to vehicle is tackled here within a Bayesian object tracking framework. However, it is emphasised that the objective in this paper is inferring the user's \textit{intent} and \textit{not} accurately estimating his/her position or velocity, as is common in classical tracking applications \cite{li2003survey,BarShalomBook2011,haug2012}. Consequently, a novel simple   prediction solution with notably low training requirements, unlike typical data-driven methods \cite{bishop2007pattern}, is proposed. It  facilitates the incorporation of contextual information such as the user's (learnt) patterns of behaviour, time of day, location, calender events, etc. Furthermore, it caters for variabilities in the driver/passenger motion \textit{en route} to the vehicle via assuming a stochastic motion model. The adopted formulation can also treat  irregularly spaced  and imprecise user location measurements via a continuous-time observations model with a random noise component. Therefore, it is a generic and considerably flexible framework. 

The proposed approach capitalises on the premise that the trajectory followed by the driver or passenger has long term underlying dependencies dictated by intent, e.g. returning to the vehicle. Accordingly, a Markov bridge, to the  endpoint (i.e. vehicle) of a known location, is built to capture these dependencies in the user's motion track. If the driver/passenger is not returning to the vehicle, no such bridging is introduced. This  postulates the addressed  inference task as a hypothesis testing problem, leading to an efficient implementation of the intent prediction procedure. It is shown here that utilising  modified Kalman filters suffices, including for the time of arrival estimation. Given the none experimental nature of this paper and the large number of possible scenarios (e.g.  car park layouts, nearby  vehicles or obstacles and others), results for two example smartphone GNSS trajectories are  presented to illustrate the usefulness and effectiveness of the introduced technique. \vspace{-3mm}
   \subsection{Paper Layout}
   The remainder of this paper is organised as follows. Related work is highlighted in Section \ref{sec:RelatedWork} and the tackled inference problem is stated in Section \ref{sec:Problem}.   The proposed Bayesian framework  and inference routine are described in Sections \ref{sec:Modelling} and \ref{sec:Inference}, respectively. Several key considerations are outlined in Section \ref{sec:PracticalConsiderations}  and the predictor performance is assessed in Section \ref{sec:Results}.   Finally, conclusions are drawn in Section \ref{sec:Conclusions}. 
\section{Related Work}\label{sec:RelatedWork}
Knowing the  destination of a tracked object (e.g. a pointing apparatus, pedestrian, vehicle, jet, etc.) can offer vital information on intent, enabling smart predictive  functionalities and automation. It has numerous application areas comprising, but not  limited to,
\begin{itemize}
\item 
 \textit{Human computer interaction HCI:} early predictions of the on-display item the user intends to select significantly reduces the interactions effort, e.g.  whilst driving \cite{AhmadMag2017}.
\item\textit{ADAS:} predicting maneuvers at intersections \cite{wiest2015probabilistic}, pedestrians motion \cite{volz2016predicting,kitani2012activity},  driver behavior \cite{bando2013unsupervised}, etc.  \item \textit{Surveillance:} inferring an object intent (e.g. ship in maritime applications \cite{piciarelli2008trajectory,pallotta2014context}) can unveil potential conflict, opportunities and facilitate automated decision making. 
\item \textit{Robotics:} intelligent  navigation in general or in the presence of other moving agents such as people \cite{huntemann2013probabilistic,chiang2015stochastic,best2015bayesian}.
\end{itemize}

Several studies in the object tracking area consider the task of incorporating  predictive, often known, information on the object's destination to improve the accuracy of estimating its state $x_{t}$  (e.g. the object's position, velocity and higher order kinematics), hence destination-aware tracking  \cite{castanon1985algorithms,baccarelli1998recursive,fanaswala2015spatiotemporal}. Furthermore, a plethora of well-established techniques for estimating $x_{t}$ from  noisy sensory observations, including the data fusion aspect, exist  \cite{li2003survey,BarShalomBook2011,haug2012}. This is referred to by conventional \textit{sensor-level tracking}. In this paper, we treat the problem of predicting the intent of a tracked object (i.e. driver or passenger)  and not estimating $x_{t}$, e.g.  his/her position. This operation belongs  to a higher system level, thus \textit{meta-tracking}, compared  with the  sensor-level algorithms. 

A destination-aware tracker with an additional  mechanism to determine the object's intended endpoint is  described in \cite{fanaswala2015spatiotemporal}. It employs   discrete  stochastic reciprocal  or  context-free grammar   processes. The state $x_{t}$ space  is discretised into predefined  regions, which the object can pass through on its journey to destination. This discretisation can be a burdensome complex task, especially if the surveyed space is large. In contrast,  in this paper we adopt  continuous state space  models with bridging distributions, which do not impose any restrictions  on the path the object has to follow to its endpoint. It is a  simple low-complexity Kalman-filtering-based solution compared with that in \cite{fanaswala2015spatiotemporal}.

Various  data driven prediction-classification methods that rely on a dynamical   model and/or \textit{pattern of life} learnt from previously recorded tracks  exist, e.g. \cite{bando2013unsupervised,kitani2012activity,huntemann2013probabilistic,chiang2015stochastic,volz2016predicting,wiest2015probabilistic}.  Whilst such techniques typically involve  substantial parameters training from complete labelled data sets (not always available)  and have high computational cost,  a state-space modelling approach  is introduced here. It  uses  known stochastic motion and measurements models,   albeit with a few unknown parameters, as is common in the object tracking area \cite{li2003survey,BarShalomBook2011,haug2012}.  We then propose effective predictors, which   are computationally efficient and require minimal training. The latter aspect is essential in the  studied automotive application since building a sufficiently large and diverse data set  of a  user approaching vehicle in a given area such as a car park, i.e. for model learning,  can be exceptionally challenging. This is due to the dynamically changing environment, e.g. other parked cars, start position of the user, followed route and  even the utilised parking space. It is distinct from set-ups where a pedestrian moves in a confined space of limited viable paths.     

Finally,   bridging-distributions-based  inference was used in  \cite{ahmad2015ICASSP,ahmad2015bayesian}, mainly for  HCI applications. It assumes that the tracked object (e.g. pointing finger in HCI) is heading to one of  $N$  possible endpoints of known locations (e.g. selectable icons on a touchscreen). Accordingly, $N$ bridges are constructed to capture  the destination influence on the object's motion. In this paper, a new  application related to intelligent vehicles is considered. Most importantly, the scenario where the driver/passenger intended destination is unknown  (i.e. not returning to vehicle)  is addressed here unlike in \cite{ahmad2015ICASSP,ahmad2015bayesian}; it is dubbed the null hypothesis. This  alters the overall problem formulation and subsequently the prediction procedure.        
\section{Problem Statement}\label{sec:Problem}
For the $n^{th}$ driver or passenger, the objective  is to  calculate the probabilities of the following two hypotheses:
\begin{align}\label{eq:HypTest}
&\mathcal{H}_{0,n}: \text{User $n$ is returning to the vehicle,} \nonumber
\\&\mathcal{H}_{1,n}: \text{User $n$ is not returning to the vehicle,} 
\end{align}
and estimate the time $T_{n}$ she/he    reaches the car, i.e. posterior $p(T_{n}|y_{1:k,n},\mathcal{H}_{1,n})$, from the available (noisy) measurements of the user's position $y_{1:k,n}$.  Observation   $y_{k,n}$  is the 2-D or 3-D coordinates of  driver/passenger at the time instant $t_{k}$, possibly relative to the vehicle. Measurements  $y_{1:k,n}=\left\{ y_{1,n},y_{2,n},...,y_{k,n}\right\}$ pertain to the sequential times  $\left\{t_{1,n},t_{2,n},...,t_{k,n} \right\}$. 
They can  be provided by the user's smartphone GNSS-based or Pedestrian Dead Reckoning (PDR)  services  and/or any other   specialised (proprietary) user-to-vehicle localisation solution. This encompasses vision-based systems and those  reliant on existing or dedicated RF technology, e.g.   from on/in-vehicle transceivers such as Bluetooth Low Energy (BLE), ultra-wideband, RFID/NFC and others  \cite{kang2015smartpdr,Sarkka2016,davidson2016survey, cheng2016localization,montanari2017study,faragher2015location,alarifi2016ultra}. This location information can be also based on a suitably equipped (smart) key-fob or any  portable device. In general, the proposed approach is agnostic to the employed user-vehicle-positioning solution and can handle  noisy irregular spaced observations; see  Section \ref{sec:Obs}.

 We assume that the location of the destination, i.e. vehicle,  is known to the inference module, for instance from the vehicle navigation system. To maintain the Gaussian nature of the formulation and for simplicity, the vehicle  is defined by the multidimensional Gaussian  distribution  $\mathcal{V}\backsim\mathcal{N}(a_v,\Sigma_v)$. Whilst the mean vector $a_v$ specifies the location/centre of the vehicle, the covariance matrix  $\Sigma_v$ (of appropriate dimension) sets its extent and orientation.

Posteriors $p(\mathcal{H}_{d,n}|y_{1:k,n})$, $d=1,2$, and $p(T_{n}|y_{1:k},\mathcal{H}_{1,n})$, are calculated at  the arrival of a new observation, hence a sequential implementation is desired.  Additionally,    computational efficiency is crucial to achieve a (near) real-time response. This is especially critical to smartphones-based implementation given the ubiquity of their location-based  services. Nevertheless, in a connected vehicle environment, computations can  be performed by the vehicle and/or cloud. It is noted that the $``n"$ subscripts  are omitted in the remainder of this paper for notation brevity.

\section{Bayesian Framework: Modelling and Bridging}\label{sec:Modelling}
Within a Bayesian formulation, we have
\begin{equation}\label{eq:Bayes}
p(\mathcal{H}_{d}|y_{1:k})\propto p(y_{1:k}|\mathcal{H}_{d})p(\mathcal{H}_{d}),~~d=1,2,
\end{equation}
where $p(\mathcal{H}_{d})$ is the prior  on whether a driver/passenger is returning to the vehicle; it is independent of the current walking track  $y_{1:k}$. This prior   can be attained from relevant contextual information $\mathbb{I}$,  such as the time of day, location of the vehicle, previous driving times, calender, etc. It can be linked to   $\mathcal{\mathbb{I}}$, i.e. $p(\mathcal{H}_{d};\mathbb{I})$. Prior $p(\mathcal{H}_{d};\mathbb{I})$ can be obtained from another system  (or even from the cloud in a connected set-up) where the user travel habits can be learnt from historical data, e.g. based on the smartphone GNSS tracks as in \cite{vlassenroot2015use,nawaz2016smart}. It  can also be gradually and dynamically learnt as the system is being used, starting from uninformative ones where both hypotheses are equally probable in equation (\ref{eq:Bayes}).

This makes the introduced framework particularly appealing as  additional information (when available) can be easily incorporated.   Therefore, the objective of the inference module becomes estimating the observation likelihoods $p(y_{1:k}|\mathcal{H}_{d})$, $d=1,2$ in (\ref{eq:Bayes}).

\subsection{Motion Models}
The driver/passenger walking motion towards the vehicle or under $\mathcal{H}_{0}$   is not deterministic. It is governed by a complex motor system and is likely to be subjected to external factors such as obstacles. Stochastic  continuous-time  models, which represent the   motion dynamics by a   continuous-time Stochastic Differential Equation (SDE), are a natural choice to suitably include the present uncertainties. This is under the premise that the intent influence on the object's motion is  captured, e.g. via bridging as in Section \ref{sec:bridging}.  Here,
 no detailed map of the environment is assumed to be available since  obstacles (e.g. other vehicles) or  moving agents (e.g. pedestrians) can dynamically change in a car park.

It is noted that the objective in this paper is not to accurately model the walking behaviour of a pedestrian. A motion model that facilities determining the probabilities of the driver/passenger returning to the vehicle suffices, however being approximate, for instance to reduce the prediction/estimation complexities. Consequently, Gaussian Linear Time Invariant (LTI) motion models are applied below as they lead to a   computationally efficient predictors, compared with  non-linear and/or non-Gaussian models \cite{cappe2007overview,BarShalomBook2011,haug2012}.
Upon integrating the  SDE,  the relationship between the system state $x_{k}$  of dimension $s\times 1$ (e.g. the drive/passenger position, velocity, etc.) at times $t_k$ and $t_{k-1}$  can be written as
\begin{equation}\label{eq:DynamicsModel}
x_{k} = F(h)x_{k-1} + M(h) + \varepsilon_k,
\end{equation}
with $\varepsilon_k\sim \mathcal{N}\left( 0,Q(h)\right) $ is the dynamic noise embodying the randomness in the motion. Matrices $F$ and $Q$ as well as  vector $M$, which together  define the state transition from one time to another, are functions of the time step $h=t_{k}-t_{k-1}$. 

The class in (\ref{eq:DynamicsModel})  encompasses  many  models used widely in tracking applications, such as the (near) Constant Velocity (CV) or constant acceleration and others   that can describe higher order  kinematics. For a CV model in 2-D, $x_{k}\in\mathbb{R}^{4}$,  $s=4$ for the position and velocity in each dimension. Models that intrinsically depend on an endpoint, such as the Linear Destination Reverting (LDR) models \cite{ahmad2015bayesian}, are covered by (\ref{eq:DynamicsModel}), for example the mean reverting diffusion model (based on an Ornstein-Uhlenbeck process), with its mean equal to $a_v$. 

In general,  models, including Gaussian LTI, which better represent the walking behaviour produce more accurate predictions as well as estimations of the  
 system state $x_{k}$. We recall that estimating $x_{k}$ is not sought here. Accurate modelling of  a pedestrian walking behaviour is currently receiving notable attention due to the growing interest in PDR and indoor positioning from smartphones sensory data \cite{kang2015smartpdr,davidson2016survey,Sarkka2016}.     \vspace{-3mm}
\subsection{Observation Model}\label{sec:Obs}
To preserve the linear Gaussian nature of the system, the   $k^{th}$ observation of the user position at $t_{k}$, is  modelled as a linear function of the  state perturbed by additive Gaussian noise
\begin{equation}\label{eq:ObsModel}
y_k = Gx_{k} + \nu_k,
\end{equation}
where $G$ is a matrix mapping from the hidden state to the observed measurement and $\nu_k \sim \mathcal{N}(0,V_k)$. For example, if the smartphone GNSS service  provides the driver/passenger 2-D position and the system state $x_k$ comprises only position, then  $G$ is a $2 \times 2$ identity matrix. The noise covariance  can be utilised to set the level of measurements noise in each axis.No assumption is made about the observation arrival times $t_k$ and irregularly spaced  measurements  can naturally be processed. 
 \vspace{-3mm} 
\subsection{Bridging Distribution}\label{sec:bridging}
For hypothesis $\mathcal{H}_{1}$, the path followed by the driver/passenger, albeit random, must end at  the intended destination at time $T$ (i.e. he/she reaches the vehicle). This can be modelled by a \textit{pseudo-observation} $\tilde y_{T}$ at $T$ or an \textit{artificial}  probability distribution for   $x_T$. This prior is equal to that of the destination $\mathcal{V}$ and its geometry modelled by $\mathcal{N}(a_v,{\Sigma_v})$. Its inclusion entails the conditioning of the motion state model  in (\ref{eq:DynamicsModel}) not only on  $\mathcal{V}$, but also on the unknown arrival time $T$.  This  permits the posterior of the system state at time $t_{k}$ to be expressed as $p(x_{k}\mid y_{1:k}, T, \mathcal{H}_{1})$, and hence the observation likelihood  $p(y_{1:k}\mid   T, \mathcal{H}_{1})$  in (\ref{eq:Bayes}). 

\begin{figure}[t]
\centering
\includegraphics[width=0.55\linewidth]{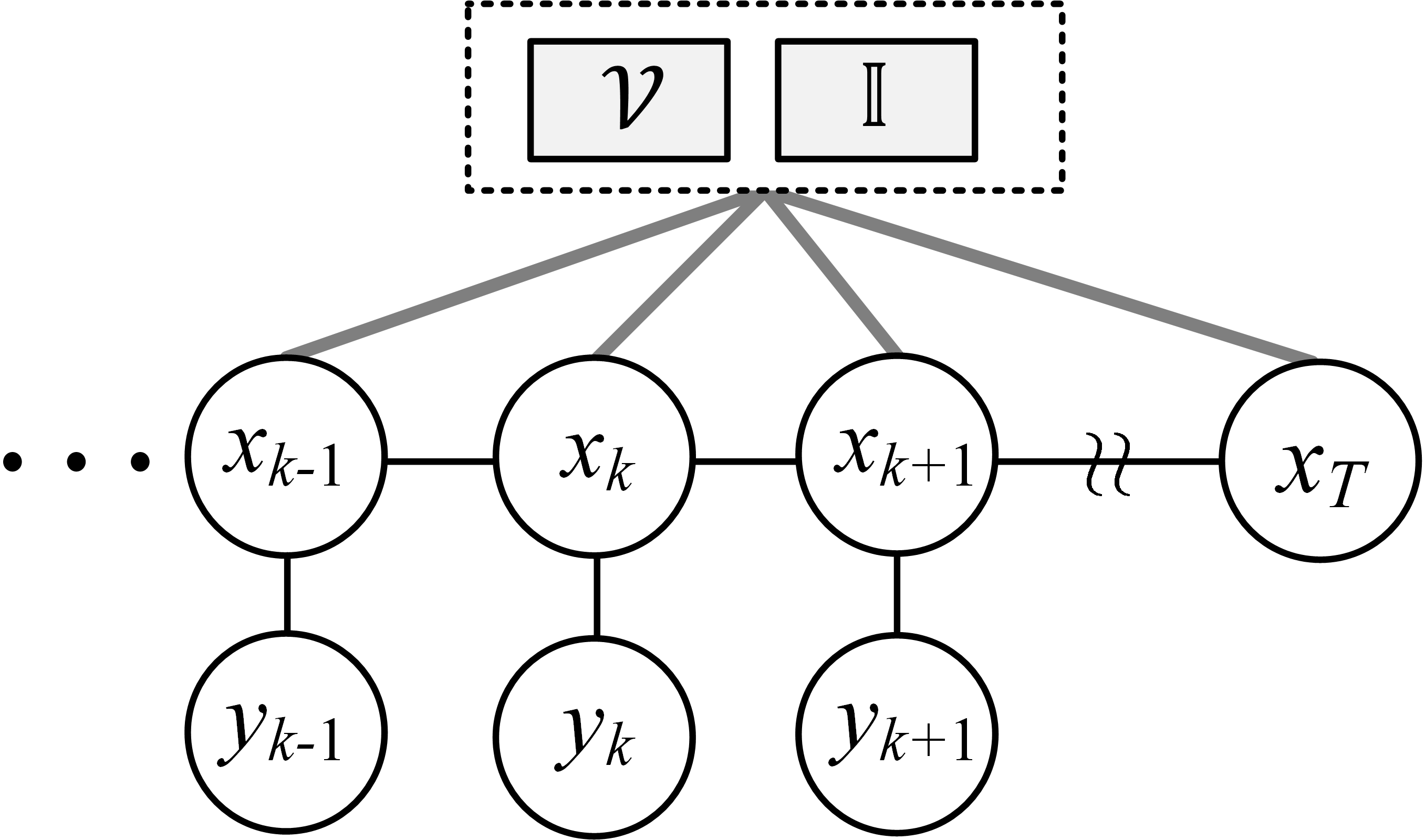}
\caption{{\footnotesize The system graphical structure under $\mathcal{H}_{1}$; destination $\mathcal{V}$  drives the state transition; contextual information, which affects intent, is also shown.  }}
\label{fig:structure}
\end{figure}

The incorporation of this destination prior changes the system dynamics, where  the predictive distribution of the user's state  changes from a random walk (i.e. with respect to the endpoint) to a bridging distribution, terminating at the vehicle. This encapsulates the long term dependencies in the walking trajectory due to  premeditated actions guided by  intent as  depicted in Figure \ref{fig:structure}, where endpoint $\mathcal{V}$ drives the state throughout the walking-to-vehicle action. In other words, it constructs a bridge between the state at  $t$ and destination at $T$. The approach of conditioning on an endpoint is dubbed \textit{bridging distributions} (BD) based inference. Thus, Gaussian linear models, whose dynamics are not dependent on the destination $\mathcal{\mathcal{V}}\backsim\mathcal{N}(a_v,\Sigma_v)$, e.g. Brownian motion (BM) and CV, can be utilised for intent  prediction within the presented Bayesian framework. On the other hand, the motion of the a driver-passenger not returning to vehicle is not influenced by the endpoint $\mathcal{V}$, and it is not bridged.  The dynamics of the walking track, which is not governed by the intent of returning to vehicle, are  \textit{approximated} directly by (\ref{eq:DynamicsModel}). 

\begin{figure}[t] 
 \centering 
 \begin{subfigure}[t]{0.975\linewidth}
\includegraphics[width=1\linewidth]{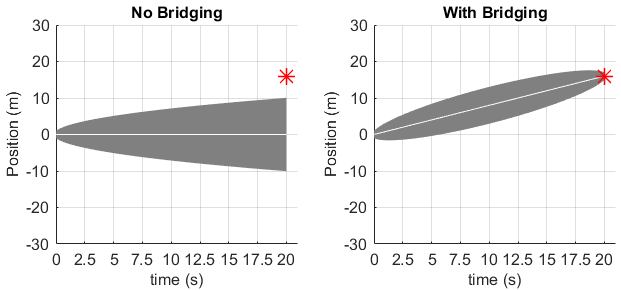}
   \caption{1-D BM model.}
   \label{fig:BMDist}
   \end{subfigure} 
 \begin{subfigure}[t]{0.975\linewidth}
 \includegraphics[width=1\linewidth]{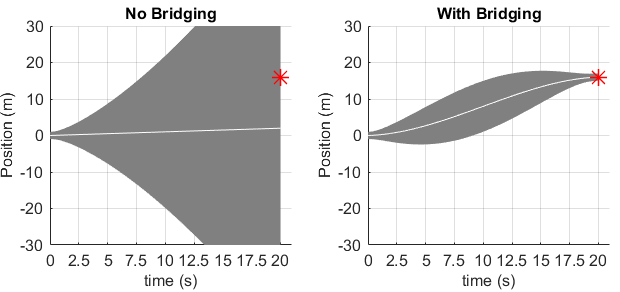}
   \caption{1-D CV model.}
   \label{fig:CVDist}
   \end{subfigure}
    \caption{Predictive distribution calculated at time $t_{1}=0$ for $t^{*}>t_{1}$ for the BM and  CV models in 1-D without (first column) and with (second column) bridging. The red star is the true destination (i.e. vehicle) located at $16$m (along the x/east axis) from the start position and reached at time  $T=20s$. Grey shading shows one standard deviation and white line is the mean;  $\sigma_{\text{CV}} = 1$ and $\sigma_{\text{BM}} = 5$.}
 \label{fig:BridgingDist}
 \end{figure}

To demonstrate the impact of incorporating  the destination prior on the motion model, the predictive position  distributions of  the BM and CV models, with and without the use of bridging, are depicted in Figure \ref{fig:BridgingDist}. The figure considers a one dimensional case where the endpoint value is $16$ m at time $T=20$s. It shows the mean and one standard deviation of the predicted position for $t^*>t_{1}$ such that $t_{1}=0$ is the current time instant.  It can be noticed in Figure \ref{fig:BridgingDist} that introducing the bridging assumption has a significant effect on the prediction results. For the non-bridged cases, the prediction uncertainty grows (arbitrarily) as $t^*$ increases. This illustrates the ability of bridging distributions to promote more accurate predictions of the intended endpoint, e.g. vehicle. 

\section{Intent Prediction}\label{sec:Inference}
 Here, we detail the means to calculate the sought observation likelihoods $ p(y_{1:k}|\mathcal{H}_{d}),~d=1,2$, in (\ref{eq:Bayes}) and time of arrival posterior. The overall inference routine  is shown in Figure \ref{fig:BlockDiagramInference}. \vspace{-3mm} 
\subsection{ Hypothesis $\mathcal{H}_{0}$: Not Returning to Vehicle}
The  observation likelihood in (\ref{eq:Bayes})  relates to the  conditional Prediction Error Decomposition (PED), which is defined by $p(y_k\mid y_{1:k-1},\mathcal{H}_{0})$, via 
\begin{align}\label{eq:LH_H0}
p(y_{1:k}\mid \mathcal{H}_{0})=&p(y_k\mid y_{1:k-1},\mathcal{H}_{0}) p(y_{1:k-1}\mid \mathcal{H}_{0}),
\end{align}
for hypothesis $\mathcal{H}_{0}$, i.e. without conditioning on  $\mathcal{V}$ (bridging). Based on (\ref{eq:DynamicsModel}) and (\ref{eq:ObsModel}), a  Kalman filter (KF) can be conveniently  utilised to calculate the PED \cite{haug2012}, recalling the Gaussian LTI nature of the motion and observation models. This is distinct from the common uses of KF in tracking application, i.e. to estimate the state $x_k$ and its posterior \cite{haug2012}. The  likelihood   $ p(y_{1:k-1}\mid \mathcal{H}_{0})$ in (\ref{eq:LH_H0}) is calculated at the previous time instant $t_{k-1}$, given the filter's recursive nature. 
\vspace{-2mm} 
\subsection{Hypothesis $\mathcal{H}_1$: Bridging Distribution}
Similar to $\mathcal{H}_{0}$,  the PED under hypothesis $\mathcal{H}_{1}$ is sought. Conditioning on the destination $\mathcal{V}$ also entails conditioning on the time of arrival $T$. One approach to introduce this conditioning is by augmenting the system state $x_{k}$ with the prior $\mathcal{N}(a_v,\Sigma_v)$ forming an extended state $z_k =[x_k' ~ x_{T}']'$ of dimension $2 s\times 1$. This can be shown to lead to the extended linear Gaussian state model 
\begin{equation}\label{eq:Zstate}
{
z_{k}=R_kz_{k-1}+ \tilde{m}_k +\gamma_k,
} \end{equation}
\cite{ahmad2015bayesian}
where $\gamma_k \sim \normdist{0}{U_k}$,  $P_{T}=\inmat{0_s & I_s}$, 
\begin{equation}\label{eq:RUm}
R_k = \inmat{H_k \\ P_T},   \quad\tilde{m}_t = \inmat{m_k\\0_s},  \quad U_k = \inmat{C_k & 0_s \\ 0_s & 0_s},
 \end{equation}
  $H_k =\left[ C_kQ^{-1}(h)F(h), ~C_kF'(_{}\tilde h)Q^{-1}(\tilde h) \right]$,   $\tilde h=T-t_{k}$, $m_k = C_k (  Q^{-1}(h)M(h)-F'(\tilde h)Q^{-1}(\tilde h)M(\tilde h))$ such that $C_k =\left( Q^{-1} (h)+ F'(\tilde h)Q^{-1}(\tilde h)F(\tilde h) \right)^{-1}$; $Q(.)$ is from equation (\ref{eq:DynamicsModel}). The observation model of dimension $k\times 1$ (e.g. $k=2$ for 2-D GNSS observations) can then be expressed by
\begin{equation}\label{eq:Zobs}
{
y_k = \tilde G z_{k}+ \nu_k
}\end{equation}
with $\tilde G = [G, ~0_{k\times s}]$ and where $G$ and $\nu_k$ are from (\ref{eq:ObsModel}).

\begin{figure}[t]
\centering
\includegraphics[width=0.85\linewidth]{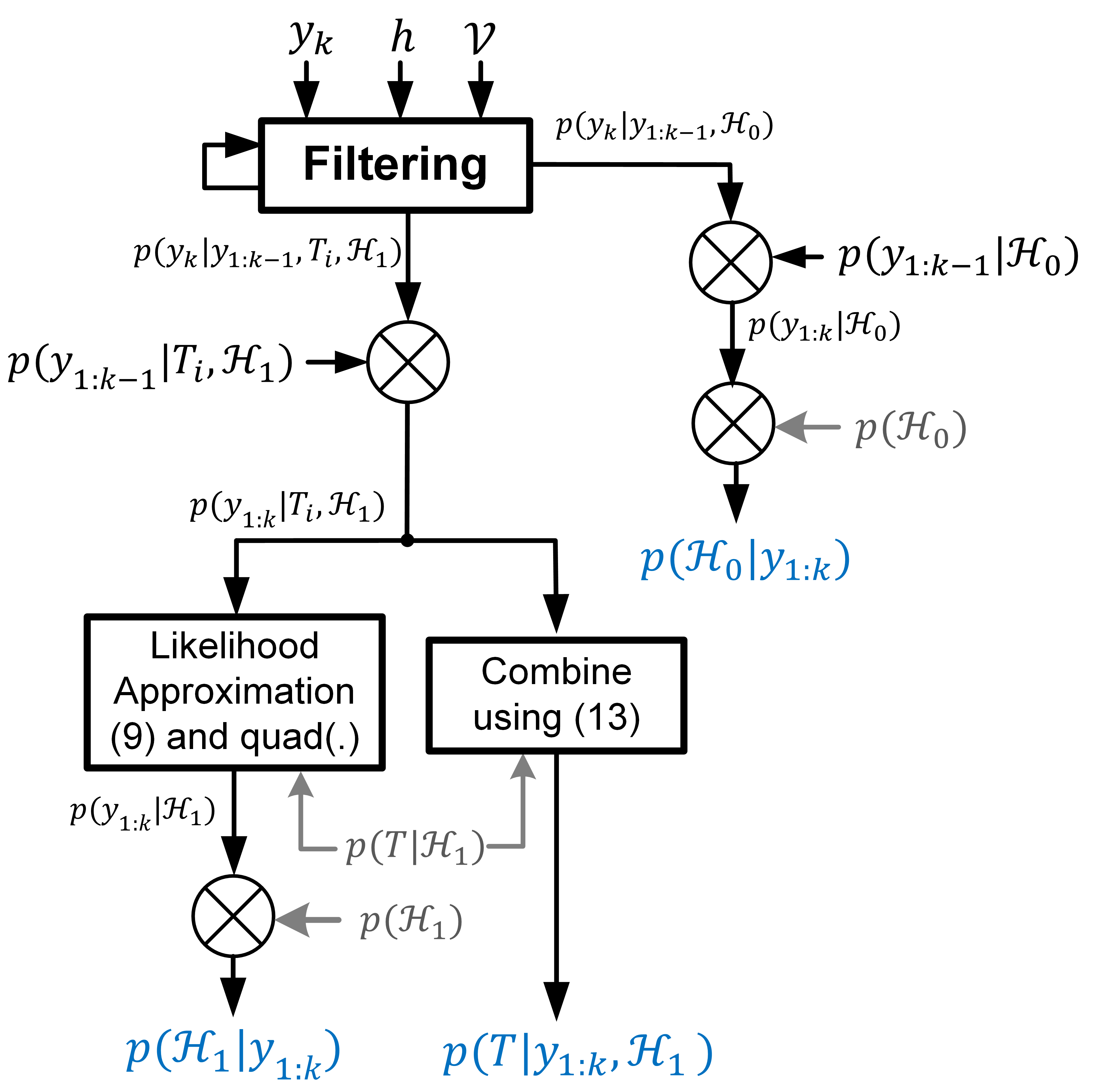}
\caption{{\footnotesize Block diagram for calculating the probabilities of the two hypotheses and estimating  $T$ after the arrival of $y_{k}$ at $t_{k}$; they are based on the Kalman filter PED. Arrow around the filtering block signifies its  recursive nature. Models parameter and priors on  state as well as   $T$ are not shown. }}
\label{fig:BlockDiagramInference}
\end{figure} 
 
Therefore, the extended system described by equations (\ref{eq:Zstate}) and (\ref{eq:Zobs})  form a Gaussian LTI system. A modified Kalman filter  can then  be applied  to  obtain the \textit{time-of-arrival--conditioned} PED defined by $p(y_k\mid y_{1:k-1},\mathcal{H}_{1},T)$ at $t_{k}$ and the likelihood $p(y_{1:k}\mid \mathcal{H}_{1},T)=p(y_k\mid y_{1:k-1},\mathcal{H}_{1},T) p(y_{1:k-1}\mid \mathcal{H}_{1},T)$ can be subsequently obtained. 

However, the arrival time is unknown in practice and typically a prior distribution on $T$ can be assumed, e.g. from contextual data. Hence, the unknown arrival time  $T$ is treated as a nuisance parameter, which must be integrated over
as per
\begin{equation}\label{eq:integral}
p(y_{1:k}\mid \mathcal{H}_{1}) = \int_{T\in\mathcal{T}} p(y_{1:k}\mid \mathcal{H}_{1},T)p(T\mid \mathcal{H}_{1}) dT,
\end{equation}
where $p(T\mid \mathcal{H}_{1})$ is the a prior distribution of arrival times and $\mathcal{T}$ is the time interval of possible arrival times $T$.  For example, arrivals might be expected uniformly within some time period $[t_a,t_b]$, giving  $p(T\mid \mathcal{H}_{1}) = \mathcal{U}(t_a,t_b)$, for instance when the user is within a certain proximity of the vehicle.

  Since the integral in (\ref{eq:integral}) cannot be easily  solved analytically for all $t_{k}$ values, a numerical approximation is applied. This  is viable since the arrival time is a one-dimensional quantity. Here, numerical quadrature   such as Simpson's rule, denoted by $\text{quad}(.)$, is utilised; other numerical methods can be employed. This approximation requires  $q$ evaluations of the arrival-time-conditioned-observation likelihood for the various arrival times $T_{i}\in\mathbb{T}$ and $\mathbb{T}=\left\{ T_{1},T_{2},...,T_{q} \right\}$, i.e. $q$ is the number of quadrature points.

\Algo{2} details the filtering procedure at  $t_{k}$ where $\ell_{k,i}=p(y_k\mid y_{1:k-1},\mathcal{H}_{1}, T=T_i)$  is the PED  for arrival time $T_{i}$ and similarly $L_{k,i}$ is the arrival-time-conditioned-observation likelihood. It is based on  Kalman filtering, notated by $\text{KF}(.)$. The filtering is performed $q$ times at $t_{k}$ to estimate $p(y_{1:k}\mid \mathcal{H}_{1})$ in equation (\ref{eq:Bayes}). This inference approach is particularly amenable to parallelisation, where the calculation of each $\ell_{k,i}$ and $L_{k,i}$ can be carried out by a  separate computational unit. This is relevant to distributed implementations in a connected vehicle environment; see Section \ref{sec:PracticalConsiderations}. 

For simplicity and at time $t_{1}$, the Kalman filtering initialisation for $\mathcal{H}_{1}$  (not shown in \Algo{2}) can be based on  
\eqn{Zprior}
{
p(z_{1}\mid T,{\mathcal{H}}_{1}) = \normpdf{\inmat{x_{t_1}\\\tilde y_{T}}}{\inmat{\mu_1\\a_v}}{\inmat{\Sigma_1 & 0_s \\0_s & \Sigma_v}},
}
where $\mu_1$ and $\Sigma_1$ specify the initial prior on $x_{1}$ \cite{ahmad2015bayesian}, thereby $p(x_{t_1}) = \normpdf{x_{1}}{\mu_1}{\Sigma_1}$;~$a_v$ and $\Sigma_v$ come from the endpoint prior $\mathcal{V}$ assuming independence between     $x_{1}$ and $\tilde y_{T}$.

After determining the probabilities of both hypotheses according to their  calculated observation likelihoods and priors, they are normalised to ensure that them sum to 1 as per:
\begin{equation}
\hat p(\mathcal{H}_{d}|y_{1:k})=\frac{p(\mathcal{H}_{d}|y_{1:k})}{p(\mathcal{H}_{0}|y_{1:k})+p(\mathcal{H}_{1}|y_{1:k})},~~ d=1,2.
\end{equation}

%
%
\begin{algorithm}
\begin{algorithmic}
\STATE \textbf{Input:} $y_{k}$, $\hat z_{k-1,i}, \Sigma_{k-1,i}$, $L_{k-1,i}$, $i=1,...,q$

\FOR{quadrature point $i\in 1,...,q$}
\STATE Calculate $R_{k}^{i}$, $U_{k}^{i}$ at  $t_k$ in (\ref{eq:RUm}), and arrival time $T_i$
\STATE Run Kalman filter:
\STATE $~~\{\ell_{k,i}, \hat z_{k,i}, \Sigma_{k,i}\}=$ 
$\text{KF}(y_k, \hat z_{k-1,i}, \Sigma_{k-1,i},R^{i}_k,U^{i}_k,\tilde G)$
 \STATE Recursive Likelihood Update:  $L_{k,i} = L_{k-1,i}\times \ell_{k,i}$
\STATE Calculate likelihood approximation:
\STATE ~~~~~~~~~$\hat L_{k,i}  =  \text{quad}(L_{k,1}, L_{k,2},..., L_{k,q})\approx p(y_{1:k}\mid \mathcal{H}_{1})$
 
\ENDFOR

\end{algorithmic}
\caption{Estimating Probability of Hypothesis $\mathcal{H}_{1}$ at $t_{k}$}
\label{alg:2}
\end{algorithm}
\vspace{-2mm}
\subsection{Estimating Time of Arrival }\label{sec:arrival-time}
The  filtering results for  inferring the  probability of hypothesis $\mathcal{H}_{1}$  in \Algo{2} can be also be readily utilised to estimate the  posterior distribution of the time of arrival of the user at the vehicle as illustrated in Figure \ref{fig:BlockDiagramInference}. It is given by
\begin{equation}\label{eq:T-posterior1}
p(T\mid \mathcal{H}_{1}, y_{1:k}) \propto p(y_{1:k}\mid T, \mathcal{H}_{1})p(T\mid \mathcal{H}_{1}),
\end{equation}
where $p(T\mid \mathcal{H}_{1})$ is the prior on the arrival time. Prior $p(T\mid \mathcal{H}_{1})$ can be also attained from contextual information including pervious journeys in a given location or   the user proximity to the vehicle as provided by a localisation module. 

The  quadrature  procedure is applied to approximate the integral in (\ref{eq:integral}) and  estimating the  likelihood $p(y_{1:k}|\mathcal{H}_{1})$ necessitates calculating the arrival-time-conditioned likelihood $p(y_{1:k}\mid T=T_i, \mathcal{H}_{1})$, for a number of quadrature points $T_i$. As a result,  a discrete approximation of the overall posterior can be obtained via  
\begin{equation} \label{eq:T-posterior-approx1}
p(T\mid \mathcal{H}_{1}, y_{1:k}) \approx \sum_{i=1}^q w_i\delta_{\{T_i\}},
\end{equation}
where $\delta_{\{T_i\}}$ is a Dirac delta located at the $i^\text{th}$ quadrature point $T_i$. To ensures that this approximate posterior distribution in (\ref{eq:T-posterior-approx1}) is a valid probability distribution that integrates to 1, it is normalised as per
\begin{equation}
w_i = \frac{p(y_{1:k}\mid T=T_i, \mathcal{H}_{1})p(T_i\mid \mathcal{H}_{1})}{\sum_{i=1}^qp(y_{1:k}\mid T=T_i, \mathcal{H}_{1})p(T_i\mid \mathcal{H}_{1})}.
\end{equation}
Thus, the 1-D posterior distribution of the arrival time  at the vehicle  can be calculated without significant further calculations beyond the already performed filtering operations. Point estimates of  $T$ can be attained, e.g. via a Maximum \textit{a Posteriori} (MAP) criterion. 
 \subsection{Decision}
 Having determined the sought probabilities $p(\mathcal{H}_{d}|y_{1:k})$,  a decision on whether the user is returning to a given entity is taken upon minimizing a cost function according to
\begin{equation}\label{eq:Cost}
 \mathcal{ \hat H}(t_{k})=\underset{ \mathcal{H}_{d}\in \mathbb{H}}{\arg\min}\;\;\mathbb{E}\left[ \mathcal{C}(\mathcal{H}_{d},\mathcal{H}^{+}) \mid\:y_{1:k}\right]
\end{equation}
where $\mathbb{H}$ is the set of considered hypotheses, e.g. $\mathbb{H}=\{\mathcal{H}_0,\mathcal{H}_1\}$, and $\mathcal{C}(\mathcal{H},\mathcal{H}^+)$ is the cost of choosing a given hypothesis $\mathcal{H}_d$ whilst $\mathcal{H}^{+}$ is the true hypothesis (intent). It can be easily seen that a binary cost function results in a MAP estimate, i.e. the most probable hypothesis is chosen. Alternatively, a threshold criterion can be used, e.g.  $p(\mathcal{H}_1|y_{1:k})\geqslant \gamma$  deems that the tracked object is returning to the vehicle; same can be applied to the not returning hypothesis. This permits quantifying the certainly level of the intent inference process and establishing cases when the system cannot determine, with acceptably high probability, the driver/passenger(s) intent.

\section{  Practical Considerations}\label{sec:PracticalConsiderations}
Here, we address the following key practical aspects of the introduced inference framework:
\begin{itemize}
\item
\textit{Computational Complexity}:
Kalman filters are known to be computationally efficient and the proposed solution has an overall computational complexity in the order of $\mathcal{O}\left( s^2 + 4qs^2 \right)$. This is a relatively low, especially given the low dimensionality of potential Gaussian LTI motion models, e.g. typically $s<10$. Additionally, a small number of quadrature points usually suffices and the bridging-based inference computational complexity can be further optimised \cite{ahmad2015bayesian,ahmad2015ICASSP}. This  can facilitate employing only one  filter for both $\mathcal{H}_0$ and $\mathcal{H}_{1}$ with an supplementary correction step, in lieu of two. 
 \begin{figure}[b]
\centering
\includegraphics[width=1\linewidth]{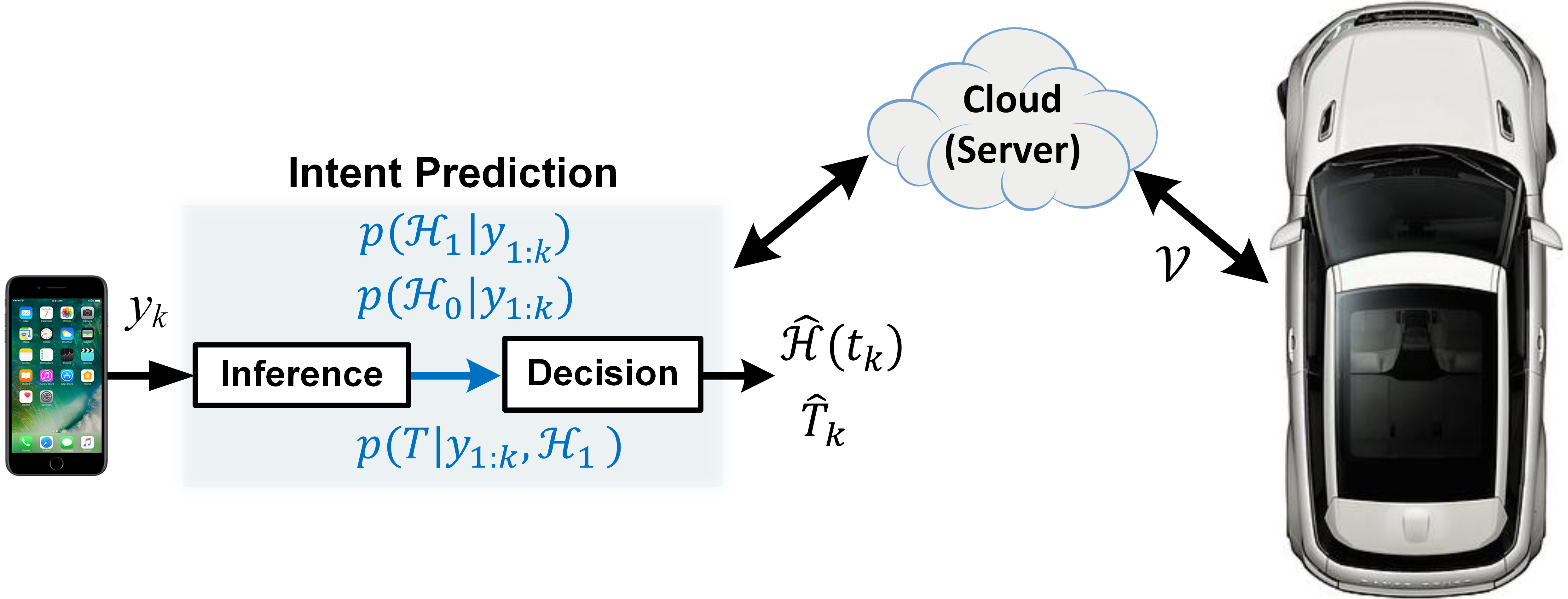}
\caption{{\footnotesize Smartphone-based implementation of the proposed solution, e.g. $y_{1:k}$ are GNSS-based position observations. The inference-decision results are shared with the vehicle via a cloud service.}}
\label{fig:SystemImp}
\end{figure} 

\item
\textit{System Implementation and Distributed Architectures}:
in a connected  vehicle environment (i.e. assuming  reliable   data links between the smartphone/portable-device, vehicle and possibly a cloud service as well as the surrounding infrastructure), the intent prediction calculations can be carried out, partially or fully, by the smartphone or vehicle or cloud. This depends on the availability of: a) the required observations $y_{1:k}$ and the vehicle information $\mathcal{V}$ (e.g. position and orientation) and b) adequate computational resources. The latter can be shared by various units within a distribution architecture given the amenability of the introduced algorithms to parallelisation; e.g. each  $\ell_{k,i}$ in Algorithm 1 can be run on a separate computational unit and all results are then aggregated. Figure \ref{fig:SystemImp} displays a possible smartphone-based implementation of the intent prediction functionality whom results are shared with the vehicle. Alternatively, the overall system can be implemented by the intelligent vehicle if  $y_{1:k}$ are locally available, e.g. from proprietary user-to-vehicle  localisation solution. Ultimately, performing the inference procedure  on the same device  providing the required measurements $y_{1:k}$, e.g. on a smartphone, minimises the communications overhead. 

\item
\textit{Training Requirements}:
  the used motion models, e.g. CV, have a notably small number of parameters; CV has only one (assuming identical motion behaviour in all spatial dimensions). These can be intuitively chosen as in the pilot results below or   based on a small number of recorded  trajectories; bridging also significantly reduces the models sensitivity to variations in the motion model parameters \cite{ahmad2015bayesian}. This clearly demonstrates the low training requirements of the approach introduced in this paper, compared with a data driven methods, e.g. \cite{kitani2012activity,huntemann2013probabilistic,volz2016predicting} where prediction models/rules are learnt from available (extensive)\ data sets. 

\item
\textit{Extensions:} this  solution is not confined to a user walking to the vehicle, other means of transport (e.g. a cart, autonomous pod, etc.) can be considered given a  representative motion model. Moreover, the prediction formulation can  be extended to  $N$ potential destinations of the driver or passenger, instead of one. All endpoints other than the vehicle will collectively constitute the null hypothesis $\mathcal{H}_{0}$. However, such a formulation involves substantially more computations as $N$ bridges need to be constructed for $N$ endpoints with their associated (filtering-inference) calculations and the numerical approximations, e.g. with $q$ quadrature points.   
\end{itemize}
\section{Pilot Results}\label{sec:Results}
\begin{figure*}[!t] 
 \centering 
 \begin{subfigure}[t]{0.396\linewidth}
\includegraphics[width=1\linewidth]{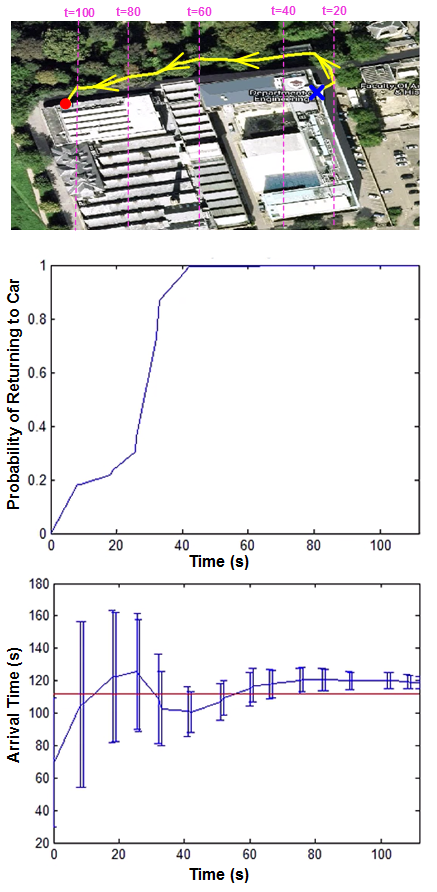}
   \caption{Returning to car.}
   \label{fig:ReturningtoCar}
   \end{subfigure}~~~~~~~
 \begin{subfigure}[t]{0.395\linewidth}
 \includegraphics[width=1\textwidth]{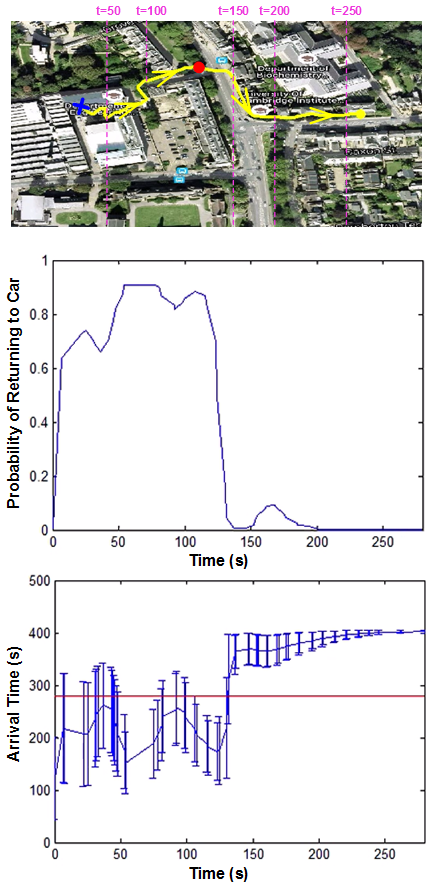}
   \caption{Walking past (not returning) to car.}
   \label{fig:NotReturningtoCar}
   \end{subfigure}
    \caption{Inference results for two trajectories for a user walking back/past the vehicle, as a function of time in seconds. First row: a map showing the followed track in yellow; arrows indicate the direction of travel, car is the red circle, start point is the blue cross and selected timestamps of the GPS trajectory are marked in pink. Second row: destination prediction as a function of time, i.e. probability that the user is returning to the vehicle $p(\mathcal{H}_{1}|y_{1:k})$. Third row: depicts the estimated time of arrival $\hat T$ (via a MAP\ estimate from posterior) as a function of time with a confidence interval (one standard deviation); red line is the true time of arrival at endpoint, i.e. vehicle only in (a).}
 \label{fig:WhetherInference}
 \end{figure*}   
Figure \ref{fig:WhetherInference} depicts the  sequentially calculated probabilities of a driver returning to the vehicle, $p(\mathcal{H}_{d}|y_{1:k}), d=1,2$, and   estimated time of arrival for two typical walking trajectories. The measurements of these 2-D tracks were collected using an Android smartphone  (assisted) GPS service at a rate of $\approx 1 \text{Hz}$. A constant velocity motion model, uniform priors on intent and time of arrivals as well as $q=40$ quadrature points are employed. A MAP criterion is utilised to obtain a point estimate of $T$ from the calculated posterior $p(T|y_{1:k},\mathcal{H}_{1})$. Figure \ref{fig:ReturningtoCar} shows results for the the scenario when a user returns to  car. Whereas, Figure \ref{fig:NotReturningtoCar} exhibits the inference outcome  when the user walks towards then past the car. Please refer to the \textit{attached video}\footnote{Alternatively, please follow the link: \textcolor{blue}{{\textit{https://youtu.be/0wHG-HqByyI}}}}  demonstrating the system response in real-time for these two trajectories.

It can be noticed from Figure \ref{fig:WhetherInference} that the proposed prediction/estimation approach provides early successful predictions in both  scenarios. For instance,  the probability of the returning-to-car $p(\mathcal{H}_{1}|y_{1:k})$  becomes significantly high early in the walking track, e.g. after $35s$ in Figure \ref{fig:ReturningtoCar}. For the second case, the inference module correctly predicts that the driver is returning to vehicle up-until the time instant $t\approx125s$, after which it  quickly changes its predictions (i.e. adapts to the situation) as the user walks past the vehicle. It is noted that the prediction algorithm has  no means of determining that the user is not returning to car, if he/she motion behaviour is consistent with walking to vehicle as in Figure \ref{fig:NotReturningtoCar}. This can be mitigated by using contextual information, such as time of day and calender. This information can be  easily incorporated within the adopted Bayesian framework via (\ref{eq:Bayes}). In terms of estimating the time of arrival at the car, the obtained results are reasonably accurate and gradually improve as more of the track becomes available. For Figure \ref{fig:NotReturningtoCar}, the  $T$  estimates deteriorate  after $t>125s$ as the user walks past the vehicle.

Conventional tracking techniques that can infer the model future state (e.g. user future position) in (\ref{eq:DynamicsModel}), i.e. without bridging, led to arbitrarily erroneous predictions similar to those in Figure \ref{fig:CVDist}.  Establishing that the driver/passenger is returning to car based on  proximity to vehicle  (e.g. when within a $10-20 \text{m}$ radius) results in late predictions and/or ambiguous incorrect decisions if the applied proximity range is increased. For instance, when the car is parked relatively  near the user's workplace or home, e.g.  within $100 \text{m}$. Contrary to these two  basic approaches, the proposed  formulation in this paper captures the intent influence on the user motion as he/she walks to car, enabling reliable early predictions of  intent and estimates of $T$. Whilst this preliminary testing illustrates the effectiveness  of the introduced approach, further evaluation from naturalistic setting is required, possibly for motion models other than the CV.

\section{Conclusions}\label{sec:Conclusions}
An simple, yet effective,  framework for predicting if and when a driver or passenger is returning to vehicle  is proposed, within a Bayesian object tracking formulation.  Notably, it is: 1)  flexible where additional contextual information can be easily incorporated, 2) adaptable where numerous motion and observation models can be used, 3) probabilistic (belief-based) where prescribed certainty requirements can be reinforced via  the decision module (or cost function), and 4) leads to low-complexity inference algorithms with minimal training requirements.
This paper sets the foundation for further work on this Bayesian approach and its applications in intelligent vehicles, including detailed experimental evaluations. 


  \section*{Acknowledgment}
 The authors would like to thank Jaguar Land Rover for funding this work under the CAPE agreement.
 \bibliographystyle{IEEEtran}
\bibliography{References}
\end{document}